\documentclass{aastex}
\usepackage{amsmath}
\usepackage{emulateapj5}
\usepackage{apjfonts}
\def\asca       {{\em ASCA}\/}
\def\chandra    {{\em Chandra}\/}
\def\rxj1720    {{ RXJ1720.1+2638}\/}

\def\rosat      {{\em ROSAT}\/}

\def\kmsmpc     {~km$\;$s$^{-1}\,$Mpc$^{-1}$}

\def\degd       {$^{\circ}\!$}
\def\second      {{\prime\prime}}

\slugcomment{ApJ in press}

\shorttitle{{\em Chandra}\/ Observation of  RXJ1720.1+2638}
\shortauthors{MAZZOTTA ET AL.}

\begin{document}

\title{{\em Chandra}\/ Observation of RXJ1720.1+2638: 
a Nearly Relaxed Cluster with a Fast Moving Core?}

\author{P. Mazzotta\altaffilmark{1}, M.  Markevitch,
 A. Vikhlinin\altaffilmark{2}, W. R. Forman,  L. P. David, and 
L. VanSpeybroeck}
\affil{Harvard-Smithsonian Center for Astrophysics, 60 Garden St.,
Cambridge, MA 02138}
\email{mazzotta@head-cfa.harvard.edu}

\altaffiltext{1}{ESA Fellow}
\altaffiltext{2}{Space Research Institute, Russian Academy of Science}

\begin{abstract}

We have analyzed the \chandra~observation of the distant ($z=0.164$) galaxy
 cluster  
\rxj1720 ~ in which we
find  sharp features in the X-ray surface brightness on opposite
sides of the X-ray peak: an edge  at 
about $250~h_{50}^{-1}$~kpc to the South-East  and a plateau at about
 $130~h_{50}^{-1}$~kpc to the North-West.
The surface brightness edge and the plateau can be modeled as a gas
 density discontinuity (jump) and a slope change (break).
The temperature profiles suggest that the jump and the break
are  the boundaries 
of a  central,  group-size
($d\approx 380h_{50}^{-1}$~kpc), dense,  
cold  ($T\approx 4$~keV) gas cloud,
embedded in a diffuse hot ($T\approx 10$~keV) intracluster 
medium.  
The density jump and the temperature change across 
the discontinuity are similar
 to the ``cold fronts'' discovered by \chandra~ in A2142 and A3667, 
and suggest
 subsonic motion of this  central gas cloud
with respect to the cluster itself.
The most natural
 explanation  is that we are observing a merger
in the very last stage
before the cluster becomes fully relaxed.
However, the data are also consistent with   
 an  alternative scenario in which  \rxj1720 ~ is the result 
of the collapse of two co-located density perturbations, the first
 a group-scale perturbation collapse 
followed by a second cluster-scale perturbation collapse
 that surrounded, but did not destroy,
the first one.
We also show that, because of the core motion, the  total mass 
 inside the cluster core, derived under the assumption of
  hydrostatic equilibrium, may 
 underestimate the true cluster mass. 
If widespread, such motion may  partially explain 
  the  discrepancy between X-ray   
and the strong lensing mass determinations
 found in  some clusters.

\end{abstract}



\section{INTRODUCTION}

In the hierarchical cosmological scenario, clusters of galaxies grow 
through gravitational infall and merging of smaller
groups and clusters (e.g. White \& Rees 1978; Davis et al. 1985).
The process may be described  as a sequence of merging and
relaxation phases in which  substructures are alternatively  accreted 
and merged into the larger system.  
To date it is not  clear to  what  extent
 subclumps of matter, that have already collapsed prior to infall into a larger cluster, can survive.
Analytic calculations, for example, suggest that groups of
galaxies may not survive  in a larger cluster environment for 
more than  one crossing time after which they will be tidally disrupted (see
e.g. Gonz\'alez-Casado, Mamon, \& 
Salvador-Sole 1994). 
However,   recent   high
resolution N-body simulations show that  galaxy-like and
group-like halos form and may survive even 
in  central cluster cores (Moore et al. 1999; Ghigna et al. 1999;
Klypin et al. 1999).

We present a \chandra ~ observation  of 
the cluster of galaxies \rxj1720 ~ that may exhibit just such a surviving
halo in its center. 
This cluster was first observed 
 in  X-rays by the  Einstein IPC 
in a project devoted to the study of the X-ray emission from radio
sources with steep spectra (Harris et al. 1988). 
It was subsequently observed with \asca ~  but, to
 date, there are no published papers on the analysis of these data,
and thus,
 we performed our own \asca ~
spectral analysis.
Among  the galaxy members only the redshift of the central
 galaxy is known
 ($z=0.164\pm 0.004$) (Harris et al. 1988; Crawford et 
al. 1999). 

The \chandra ~ data  show a sharp edge in the X-ray surface brightness
produced by a density discontinuity at 
about $250~h_{50}^{-1}$~kpc  SW  of the X-ray peak.
The density jump and the temperature change across this discontinuity
 are similar
 to the edges observed by \chandra~ in A2142 (Markevitch et al. 2000b) and 
A3667 (Vikhlinin, Markevitch, \& Murray 2000a).
On the opposite side, at about  $130~h_{50}^{-1}$~kpc 
 NE of the X-ray peak,  
the surface brightness shows a plateau which is consistent with a
 break in the slope of the radial gas density distribution. 
We show that the X-ray features are
 consistent   with being
the boundaries   of a group-size, 
 dense gas cloud moving 
with respect to the cluster.
In $\S 3.1$ we discuss the possibility that, as already suggested for 
A2142 and A3667, \rxj1720 ~ is a 
  merger.  
In $\S 3.2$ we  explore a possible dual potential structure for
\rxj1720 ~  resulting from  
 a group-size density perturbation collapse 
followed by a second cluster-scale perturbation collapse at nearly the same 
location in space.

Regardless of the cluster history, in $\S 3.3$ we
 show that, 
 because of the cloud motion,
the cluster mass obtained using
the hydrostatic  equilibrium equation, on scales 
smaller than the gas cloud size, 
may be an underestimate.
If \rxj1720 ~ is not a ``special''
 cluster and the core  motion is
present in many other clusters, then it may partially explain 
  the  discrepancy between X-ray   
and the strong lensing mass determinations
 found in  some systems.

We use  $H_0=50h_{50}$\kmsmpc and $\Omega_{0}=1$, so at  the
cluster redshift of $z=0.164$, $1^\second = 3.652$ kpc; if not specified
differently, we use one-parameter
$90\%$ ($\Delta\chi^2=2.71$)   confidence intervals.


\section{DATA ANALYSIS}

\rxj1720 ~ was observed by \chandra\  on  1999
October 19 in  ACIS-I
\footnote{\chandra\ Observatory Guide
http://asc.harvard.edu/udocs/docs/docs.html,
section ``Observatory Guide'', ``ACIS''.}
for a useful exposure time of 7.57ks.
Hot pixels, bad columns, chip node boundaries, and events with
 grades 1, 5, and 7 where excluded from the analysis. The cluster was
nearly centered in the I3 chip.
To exclude from the observation any period with anomalous background, 
we made a light curve for each ACIS-I chip and checked that the
background did not exhibit any strong variations on time scales 
of a few hundred seconds.  

Although the particle background was rather constant in
time, it is non-uniform over the chip (varying by $\sim 30$\%). 
To take this nonuniformity into account in our spectral and
imaging analysis,
 we used the public background dataset  composed of several 
observations of relatively empty fields. These
observations were screened in exactly the same manner as the cluster data.
The total exposure of the background dataset varies from chip to chip and
it is  about 300-400ks. 
The
background spectra and images were  normalized by the ratio of the
respective exposures.  This procedure yields a background which is accurate
to $\sim 10$\% based on comparison to other fields; this uncertainty is
taken into account in our analysis (see Markevitch et al. 2000a for a 
description of the ACIS background modeling).


\subsection{Cluster Morphology}

To study the X-ray morphology of the cluster we generated 
an image with $1^\second \times 1^\second$ pixels 
from the events in the chips I0, I1, I2, and I3.
We extracted the image
in the 0.5-5~keV band  to minimize the relative contribution of the
cosmic background and 
thereby to
maximize the
signal-to-noise ratio.  
In  Fig.~1 we  overlay the ACIS-I, X-ray
contours  
 (after an adaptive smoothing of the image with a circular top hat
filter with  a minimum of 30 counts under the
filter) on the DSS
optical image.
The X-ray brightness peak coincides with the
 optical center of the cluster central galaxy.
Moreover the cluster appears spherically symmetric at  large radii and the
 centroids of the X-ray surface brightness for $r \ge 100^\second$ 
are coincident within $4^\second$ with the
  X-ray  peak.
However, 
we cannot fit a simple   $\beta$ model to the X-ray
surface brightness. Even
 excluding  the  inner
regions  $r \le 100 - 150 h_{50}^{-1} $~kpc 
($r \le 27^\second - 41^\second$) to account for an apparent 
 cooling flow, we find a reduced $\chi^2>3$.
The failure of the $\beta$ model is  mainly due  to data points at
radii between
$30^\second$ and $80^\second$.
 
The isointensity contours, in fact, show that the X-ray surface brightness 
has an azimuthally symmetric 
distribution at large radii ($r\ge 100^\second$)  as well as at
 small radii ($r\le 16^\second$).  
However, there is a steep gradient in the surface brightness
 followed by a plateau  at 
$r=30^\second-50^\second$ to the North-West and a sharp edge 
at $r=70^\second$ to the South-East   of the
X-ray peak.
The SE edge remains sharp within a
 sector of angular extent $\psi\approx \pm 30$\degd ~ before 
gradually vanishing.


\subsection {Density Profiles}
To  derive the X-ray surface brightness profile across these two features,
we divided the cluster into two sectors
 centered on the X-ray peak as shown in Fig.~1: North-West
(NW) and South-East (SE). The
sector angles are -80\degd ~ to 10\degd ~ and 109\degd ~to 169\degd ~  for the 
NW and the SE regions respectively (the position angles are measured
from North toward East). We extracted a surface brightness profile
from these two sectors in circular regions as shown 
in Fig.~2a.
The SE radial profile clearly shows a sharp
edge at about $r=70^\second$, while
the NW profile shows a plateau in the region between
$r=30^\second - 50^\second$.  
 The radial derivative of the SE surface brightness profile is clearly discontinuous on a scale
 $\le 10^\second$.

The NW brightness profile  appears to be more
continuous but  it still could not be reasonably fitted 
(reduced $\chi^2 >3$) with  
the sum of two beta models. This indicates that we are not observing the
projection
 of two clusters/groups of galaxies along the line of sight.
The particular shape of the brightness profiles, 
instead, may indicate  discontinuities in the gas density
profile. 

\bigskip

{\footnotesize
{\centerline {\bf TABLE 1}}
{\centerline {Model Fits}}
\noindent
\begin{tabular}{c c c c c c c c}
\hline \hline
Sector  &  $\alpha$ & $\beta$ & $r_c$ & $A_{jump}$ &  $r_{jump}$ &
$\xi$ &$\chi^2/$ \\
 &  & & (arcsec) & & (arcsec) &(\degd)&d.o.f. \\
\hline 
NW &$2.2 $& $0.63$ & $43$  &  $0.95 $ & $35 $
&---& 48.6 \\
   &$ \pm 0.02$&  $\pm 0.02$ & $\pm 2 $& $ \pm0.06$ & $ \pm 3$
& & 46 \\
& & & & &\\
SE &$0.59 $& $0.63 $&$76 $ & $2.8 $ & $69 $
&12 & 39.5 \\
 &$\pm 0.07$& $\pm 0.04$& $\pm 11$& $\pm 0.2$ & $\pm 2$
&$^{+23}_{-47}$ &46 \\
\hline
\end{tabular}
\\
}
\bigskip

To quantify the discontinuities, we fit the  brightness
profiles with the density model defined in the Appendix. 
The idea behind the model is that we  have two
 concentric  regions with  different radial gas density distributions.
We assume that the gas density distribution is a spherically symmetric
power-law for the innermost region ($r<r_{jump}$) and the usual
$\beta-$model for the external one ($r>r_{jump}$).
The density distribution is characterized by
a discontinuity (``jump'') at $r_{jump}$
of amplitude $A_{jump}$.
If  $A_{jump}>1$, the projection in the plane of the sky of such a density
 model,  produces
an azimuthally symmetric brightness jump at $r_{jump}$.

As one can see from Fig.~1, the SE surface brightness jump is not  
azimuthally symmetric: it becomes less abrupt at some angle $\psi$.
This may indicate that the discontinuity in the
actual cluster gas distribution is smeared out at some angle.

To reproduce the observed SE brightness edge we modify the above
density model introducing, at the interface of the internal and
external regions, a boundary
layer in which the gas density changes sharply but continuously from its
value in the inner region to the value in the outer region. 
The thickness of this layer depends on the angle $\zeta$ from 
the axis of a chosen direction $\hat \delta$
(which will late define the direction of motion of the central gas
cloud), being zero at the direction of the axis (where the density profile is
truly discontinuous) and growing with $\zeta$. The physical significance of
this model boundary layer will be touched upon in the Discussion. 
The particular
functional shape of the density profile inside the layer ("matching
function" defined in the  Appendix) is not very important; we assume it is a
power law determined by the layer thickness at each given direction.
The dependence of the layer thickness on the
angle $\zeta$ can be derived by matching the two-dimensional projection of the
model to the image of the SE edge.  Once this dependence is fixed (see
Appendix), the radial brightness profile can be used to fit the remaining
parameters of the model.
We will see that the brightness profile even
allows us to constrain the direction $\hat \delta$.
In fact, the model is fit to the brightness profile with
  the free parameters being the inner 
power law slope, the core radius and the
beta-model slope ($\beta$), the angle $\xi$ of the direction
$\hat\delta$ with respect to the plane of the sky,
and the radius and the  amplitude of the jump. 
The best-fit
values within their $90\%$ errors are reported in Table~1. 
The best-fit 
 density
model is shown in Fig.~2b and the 
corresponding brightness profile
is overlaid as a histogram on the data points in Fig.~2a.
We find that the best-fit 
density  jump factor 
is $2.8$  while $\xi=12^{+23}_{-12}$~deg.
Because the density model is  symmetric
 for $\xi \rightarrow -\xi$, 
the direction  $\hat \delta$ 
should lie within $35$\degd ~ of the plane of the sky.

As shown in the Appendix (Fig.7), the constraint $|\hat \delta|<35$\degd~ 
comes from the fact that
the density edge is present in a finite sector and would not be seen
in projection if the angle were too large.

Unlike the SE sector, the NW sector  is consistent with no density
jump. Therefore, to fit the NW profile, we omit the  matching function
 from the previous density model, by  setting $L(\zeta)=0$.
The best-fit density profile parameters (within their 90\%
errors)  are given in Table~1 and the best-fit density model and the
corresponding brightness profile are shown 
in Fig.~2b and Fig.~2a, respectively.
Even if the NW density profile is consistent with no density jump, it
shows  an evident  break in slope.

Finally, we notice that, 
while the density  slopes at larger radii are similar for both the NW
and the SE regions, the slopes of the
inner regions are higher and lower for the NW and  SE sectors, respectively.

We would like to remark that the SE and NW sectors have
different density profiles in the regions we have modeled (beyond
$r=16^\second$). Thus, our model cannot be extrapolated to smaller
radii where the projected surface brightness distributions must
ultimately agree at the center even though we accurately describe the
gas density distribution at $r>16^\second$.
We also ignored the gas temperature variation which
represents a small correction ($\le 0.5\%$) for the energy band
that we are using. 

The temperature structure of these features is discussed in \S 2.4 below.


\subsection{Average Cluster Spectrum}

Spectra
were extracted in the 0.6-10~keV band in PI channels 
that correct for the gain difference
between the different regions of the CCDs. The spectra where then
grouped to have a minimum of 50 counts per bin and fitted using the
XSPEC package (Arnaud 1996). Both the redistribution matrix (RMF) and the
effective area file (ARF) for all the CCDs are  position dependent. In our
spectral analysis we computed  position dependent 
RMFs and ARFs using the CIAO 1.1.3 package
 weighted them
by the X-ray brightness over the corresponding image region.
As a result of  CTI, the quantum efficiency (QE) of the ACIS 
front-illuminated CCDs
decreases far from the read-out at high energies. 
In building the position dependent ARFs we used a  preliminary version
of the QE non-uniformity calibration file that 
corrects for this effect assuming that all chip nodes behave 
identically (Vikhlinin 2000).

We extracted an overall spectrum for  \rxj1720 ~ in the ACIS-I3 chip
 and  fitted it with an absorbed 
 single-temperature thermal model (Raymond \& Smith 1977, 1992 revision).
At the present stage of the \chandra ~ calibration the responses of the
ACIS-I chips are  uncertain below $E\approx 1$~keV. Thus  we 
initially restricted
our fit to the   1-10~keV energy band and 
we fixed
 the  equivalent hydrogen column to the Galactic value 
($N_H=4.06\times 10^{20}$cm$^{-2}$).
The result of the fit is given in Table~2.
The same table gives the result of our fit to the
1.5-10~keV spectrum
extracted from the innermost $r=4.5^\prime$  region obtained with 
the ASCA-GIS detector fixing the  equivalent hydrogen column to 
 the Galactic value.
We find that the \asca ~ best fit  temperature  is  in good agreement 
 with the \chandra ~ result.

To test how the \chandra ~  low energy calibration uncertainties 
affect our results, we refitted the spectrum 
using the  0.6-10~keV energy band and $N_H$ as a free parameter.
In this case we find that the temperature is  in excellent 
agreement with our previous result (see Table~2) but 
 the \chandra ~ data require a significantly higher 
equivalent hydrogen column.
A similar calibration effect
is seen in the \chandra ~
observation of  the Coma cluster (OBSID=1113).
We measured the temperatures by extracting 0.6-10~keV spectra 
in four rectangular regions on the
ACIS-I3 chip parallel to the read out.
We find that while the temperatures are   
within the errors of those measured by \asca ~ (Donnelly et al. 1999) the
measured absorption in each region is significantly higher than the Galactic
 value. 
We conclude that, at the present stage of the \chandra ~ calibration,
to obtain a  reliable temperature measurement it is prudent
 to restrict
the analysis to energies  1-10~keV and fix the $N_H$ value. 
However, the measured temperature is not significantly affected
when we use  the entire 0.6-10~keV energy band with  $N_H$ 
as a free parameter.

The total unabsorbed cluster flux in the  0.1-2.4~keV  energy band, measured with 
ACIS, is $f_X=14.2
\times 10^{-12}$~erg~cm$^{-2}$~s$^{-1}$, which 
 is consistent with the \rosat ~ value ($f_X=14.3
\times 10^{-12}$~erg~cm$^{-2}$~s$^{-1}$, Ebeling et al. 1998).
This flux corresponds to a 
luminosity  of $L_X=16.5\times 10^{44}$ erg~s$^{-1}$. 
From the
$L_X-T$ relation in the 0.1-2.4~keV band 
(see e.g., Markevitch 1998), we find that the
luminosity of \rxj1720 ~ is typical of a 
 $T\approx 10$~keV cluster, a temperature much higher than 
what we measured.
Below we investigate the spatial temperature structure which may be
responsible for this discrepancy.

\bigskip

{\footnotesize
{\centerline {\bf TABLE 2}}
{\centerline {Average Spectrum}}
\noindent
\begin{tabular}{l c c c c c c }
\hline \hline

   &  $T_e$ & $N_H$                 & $Z$         &  $\chi^2/d.o.f$ \\
               &    keV & $10^{20}$cm$^{-2}$  & $Z_\odot$   &  \\
\hline 
AXAF\tablenotemark{a} & $5.6 \pm 0.5$ & $4 $    &$0.44\pm 0.15$ &144.4/122\\
AXAF\tablenotemark{b} & $5.2 \pm 0.3$ & $10 \pm 1.1 $    &$0.58\pm 0.15$ &185.6/154\\
ASCA & $5.6 \pm 0.5$   &  $4  $    &$0.2\pm 0.15$ &85.5/77\\

\hline
\end{tabular} \\
$^{\rm{a}}$ [1-10]~ keV energy band \\
$^{\rm{b}}$ [0.6-10]~keV energy band \\
}


\subsection {Temperature Profile}

We first derive  a radial temperature profile of the entire cluster 
in eight annular regions centered on the X-ray peak.
Following the procedure described in the previous section, we
 extracted the overall spectrum for each region.
To quantify the corresponding uncertainties due to the low energy 
calibration problems, we fit the energy band from 
1-10~keV with $N_H$ fixed at the Galactic value
and the 0.6-10~keV band with $N_H$ free.
In  Fig.~3 we show 
the projected emission-weighted  temperature profile. 
The solid and dotted crosses show the temperature profiles obtained from the
1-10~keV and 0.6-10~keV fits respectively; they are consistent within their 
statistical errors.

The temperature profile  clearly shows  the presence
 of
at least two components: a more or less isothermal core region
($r<180h^{-1}_{50}$~kpc) with a temperature of $\approx 5$~keV,
surrounded 
by a hotter, more extended region with $T\approx 10$~ keV.

We notice that the temperature estimated from 
the temperature-luminosity relation is in 
agreement with the  temperature measurement for the
external region, suggesting
that, on average,  the properties of the 
external regions of the cluster  are those of a hot,
 luminous cluster.

We also  determined the 3-D temperature in the the innermost
 spherical $r=35^\second$ region by fitting the spectrum with a two temperature
 Raymond-Smith model, fixing the normalization and the temperature for the
hotter component to those expected from spherical projection. We find  that 
 the true temperature of this central region is $\approx 3-4$~keV.

Because of the low energy calibration uncertainties 
it was not possible to confirm or  exclude the spectroscopic signature 
of a  cooling flow in the core of \rxj1720 .
However, from  the observed  3-D temperature and 
the gas density,  we find the  cooling
time at the sharp borders of the core to be $t_c\approx 5\times 10^{9}$~Yr
(see e.g., Sarazin 1988). This time is  larger than 
the expected age of the cluster, thus it appears unlikely that the observed low
temperature of the central region is the result of a cooling flow.
Indeed, a rather constant temperature profile with a dip at the
smallest radial bin suggests that cooling is significant only at that
small radius.

To derive the detailed temperature structure across the brightness edges,
we  also  extracted  separate temperature profiles from the NW and the SE sectors,
 shown in Fig.~4.
 Due to the short exposure time,
 the statistical errors are rather large, but
   both profiles suggest  a temperature rise as one
 moves outward across the brightness edges.


\section{DISCUSSION}

Density discontinuities similar to that seen in the SE  sector 
of    \rxj1720 ~
are theoretically expected in  shock fronts in a gas flow (see Landau \&
Lifshitz 1959).
However the temperature jump  across the edge
goes in the direction opposite to that expected for a shock front.
Indeed, if we apply the Rankine-Hugoniot shock jump condition (see Landau \&
Lifshitz 1959,\S\S 82-85), a factor of $\approx 2.6$ density jump with
a post shock temperature of $\approx 3-4$~keV  (the inner region of
the SE edge), one would expect to find a $T\approx 1.5$~keV gas in
front of the shock (on the side of the edge away from the cluster
center) which is inconsistent with the much higher observed temperature 
(see Fig.~4).

Instead the observed sharpness and the temperature change of the SE edge are
  similar to those  observed with \chandra ~ in  two 
clusters with major mergers: A2142 (Markevitch et al. 2000b) and A3667
(Vikhlinin et al. 2000a).
As suggested for those clusters, the most natural 
explanation is  that the observed phenomenon 
is produced  
by the motion from NW
to SE of a central, colder,  denser body of gas  through the hotter, 
cluster ICM.
The X-ray edge is the surface of the contact discontinuity where the
pressure in the dense core is in balance with the thermal plus ram
pressure of the surrounding gas:
the outer region of the moving gas cloud where the pressure was
  lower is stripped. The edge is sharp in the direction of motion,
  but is gradually destroyed by gas dynamic instabilities at an angle
  to that direction, as in A3667 (Vikhlinin et al. 2000a).
Following Vikhlinin et al. (2000a), we
 estimate  the speed of the  dense cloud
 using the equations for 
flow past finite bodies (see Landau \&
Lifshitz 1959,\S 114):

\begin{equation}
{\frac {p_0}{p_1}}=
\left\{
\begin{array}{*{2}{l}}
\left ( \frac{\gamma+1}{2} \right )^{(\gamma +1)/( \gamma-1)} \left [
\gamma - \frac {\gamma-1}{2M_1^2} \right ]^{-{1/(\gamma-1) }}M_1^2; &
M_1>1 ; \\
\left (1+ \frac{\gamma-1} {2} M_1^2 \right )^{\gamma / (\gamma-1)} ;&
M_1\le 1.\\ 
\end{array}
\right . 
\end{equation}
\noindent
Here  $p_0$ is the pressure of  the surrounding 
gas where  the 
flow speed is zero (at the stagnation point immediately near the tip
of the body). 
If we assume that the gas in the subclump does not expand, then 
the pressure $p_0$ 
is equal to the pressure just inside the edge.
The parameters $p_1$ and  $M_1$ are  the the pressure of the incident gas 
and the Mach number (the ratio of the fluid velocity to the velocity
of sound), respectively, calculated far  from the body.
Finally $\gamma$ is the ratio of the specific heats.
Applying equation (1) to the measured pressure jump for the SE edge, we find
$M_1=0.4^{+0.7}_{-0.4}$ which suggests that the central gas cloud
moves subsonically (or at least not very supersonically)
 with respect to the surrounding gas.

Before discussing the nature of the moving core, we summarize
 the observational evidence:

\begin{enumerate} 
\item the surface brightness is azimuthally symmetric at large radii
and does not show the characteristic
irregular structure typical of major mergers;
\item  the external slopes ($\beta=0.63$)
 of the gas density are  similar 
for  the NW and the SE sectors,
proving that the gas  follows the same  spherically
symmetric  potential at large radii;

\item the total cluster luminosity and the external  temperature are
consistent with the L-T relation (see \S 2.4);

\item  the available DSS optical data, although not  deep,
exhibit only one central dominant
galaxy and it is  coincident with the  X-ray surface brightness peak;

\item the velocity  of the subclump  is small
compared to that of a point mass free-falling from infinity 
into the cluster center (assuming an isothermal cluster, $M_1\approx
2.7$; see e.g. Sarazin, 1988).

\end {enumerate}

Below we discuss two possible evolutionary scenarios.

\subsection {Major Merger}

As suggested for A2142 and A3667,
the most natural explanation for the presence of a moving subcluster
is that
 \rxj1720 ~ is
undergoing a merger.
However, if this is the first or second passage of the subclump through
the main cluster, then
most of  the  above properties suggest that  the merger direction  is
close to the line of sight. 
This is inconsistent with the presence of the SE surface brightness
edge that clearly indicates that the merger 
direction should be within  $35$\degd~
of the plane of sky (see \S~2.3).

On the other hand, if the merger direction is close to the plane of
the sky, 
then the azimuthal symmetry and the 
low speed of the subclump require that
 the two merger objects, through  damped oscillation,  have already
passed through each other several times and  are now 
 in the final stage before the cluster becomes
fully relaxed. This  raises the question of how the subclump 
could have survived   such 
multiple core crossings without   being disrupted 
by  tidal forces.
It is not  clear to  what  extent
subclumps of matter, that have already collapsed prior 
to infall into a larger cluster, can survive the infall.
Analytic calculations suggest that, because of the tidal
forces, groups of
galaxies may not survive  in a larger cluster environment for 
more than  one crossing time (see
e.g. Gonz\'alez-casado, Mamon, \& 
Salvador-Sole 1994). 
However   recent, high
resolution N-body simulations show that  galaxy-like and
group-like halos form and may survive even 
in the very central cluster region (Moore et al. 1999; Ghigna et al. 1999;
Klypin et al. 1999).
In any case it is clear that to survive tidal forces the central
subclump must have a particularly compact, deep, gravitational 
potential  that would have formed at 
high redshift.


\subsection{Collapse of Two  Density Perturbations}

As noted above, the X-ray observations suggest that
 \rxj1720 ~ is composed of an 
extended hotter gas  component with a  spherically symmetric
density distribution at radii $r > 200- 300 h_{50}^{-1}$~kpc that hosts 
a moving, apparently thermally isolated, colder,  compact ($r\approx 180 h_{50}^{-1}$~kpc)
 object near its center.

An alternative to the merger hypothesis is that
this
cluster may be the 
 result of the collapse of two different  perturbations 
in the primordial density field
on two different linear scales at nearly the same location in space.
As the density field evolves, both perturbations start to collapse. 
 The small scale perturbation collapses first 
and forms a central group of galaxies while the larger perturbation
continues to evolve to form 
 a more extended
  cluster potential.
The central group of galaxies could have formed  slightly offset
 from the center of the cluster and is now
 falling into  or oscillating around the
minimum of the potential well. This motion is responsible for
the observed surface brightness discontinuity.

A possible problem with such a scenario is that 
 we might expect  some of the mass from the
larger cluster perturbation to collapse onto the already
 virialized group by  secondary infall of  shells just beyond the
 group  (e.g. Hoffman 1988).  
If the mass accretion of the group from secondary infall is too large, then
it is plausible that the final stage of the evolution will result in
  just one single cluster potential, thus
 destroying the initial dual potential structure.
However, 
we may expect that the effect of 
 secondary infall is
small enough  that  the group may
maintain its own identity. 
In fact,  a group  is just a rescaled version 
of a galaxy (Moore et al. 1999) and  N-body simulations of the formation 
and evolution of galactic halos in
clusters clearly show  that
the mass of galactic halos that form inside a larger  cluster-size
perturbation
stops evolving well before the cluster virializes (Okamoto \& Habe
1999).
We also note that  dual 
potential structures have been suggested in previous works 
for several clusters  (Thomas, Fabian, \& Nulsen 1987; 
  Nulsen \& B\"ohringer 1995; Ikebe et al.1996).
These authors suggest that the X-ray data for these 
 clusters  are consistent with two distinct components: 
 a large-scale cluster component and a central
 compact component associated with the cD galaxy. 
This model then may represent a rescaled version of 
these previous suggestions.

Unlike a major merger, this last
interpretation does not have a problem of tidal force
disruption of the smaller central subclump. Indeed,
the speed of the subclump may be subsonic just because 
its initial position lies well within  the  main cluster.
This may indicate  that we are  observing the subclump's first or second
passage through the main cluster core and there has not been enough
time for it to be tidally disrupted.

Our model suggests that the central group 
 virialized  earlier
than the cluster itself. 
We can estimate the formation epoch of the group assuming that 
the  external, higher gas temperature reflects the depth of the
larger scale perturbation while  the internal temperature
is that of the originally colder central group, after some 
 compression
 induced by the pressure of the cluster gas.
Because the present pressure at the center of the group is much higher
than the pressure of the cluster near the group, such compression
cannot have been very significant.

Therefore, we simply assume  that the observed internal temperature 
 is  that of the  central group at the time it virialized.
We can also estimate
 the gas mass associated with the central group assuming that
the total gas mass is the sum of the gas masses inside 
the two hemispheres  
of radius $r_{jump}$ with density
profiles  given by the best-fit values in Table~1
 for both  NW and SE sectors.
We find  
$M_g= 
 8.8\times 10^{12} h_{50}^{-5/2}M_{\odot}$.
Because the central region is an ellipsoid rather than a sphere, 
our 
calculation may overestimate the actual gas mass by not more than
a factor $1/e^2$, where
$e$ is the eccentricity.
 The eccentricity
of the central region that we measured
from the X-ray surface brightness appears to be
 $\le 0.77$.
Thus, we can safely constrain the total gas mass
associated with the central group 
to be $M_g\approx 5- 
 9\times 10^{12} h_{50}^{-5/2}M_{\odot}$.

Using the $M_g-T$ relation (see e.g. Mohr et al. 1999;
Nevalainen, Markevitch, \& Forman 2000; and references therein)
and assuming that  $z_g$ is the redshift at which the group virialized,
 we find that the measured gas mass corresponds to that of a group
with a temperature of
 $T\approx (1 - 1.5) (1+z_g)$~keV.
In $\S 2.4$ we estimate the group  3-D temperature to be $T\approx
3-4$~keV.
Thus, the latter equation allows us to constrain the group 
formation redshift to  $1<z_g<3$.
From the dissipationless uniform spherical collapse model (Peebles 1980)
we know that the time $t_c$ at which a perturbation collapses and
the time $t_m$ at which the perturbation reaches its maximum expansion
are related as $t_c=2 \times t_m$.
If we assume that 
the larger scale perturbation collapsed just at the cluster
redshift ($z=0.164$), then the same perturbation
 reached its turn around point at 
$z=0.85$. Thus, at the time when the group had already formed,
 the larger scale perturbation was still
expanding with the Hubble flow and, hence, the central group was only
weakly affected by its environment, a situation in accord with our
model.


\subsection{Hydrostatic Equilibrium at the Subclump Boundaries}

Regardless of the origin of the central subcluster, it is interesting to
consider its effect on the estimate of the total mass.
For gas in hydrostatic
equilibrium,  the hydrostatic equation can be used to
 estimate the cluster mass inside a  radius $r$ 
(see e.g., Bahcall \& Sarazin 1977; Mathews 1978). 
In Fig.~5 we show the total mass profiles derived 
from the gas density properties of the SE and NW sectors, assuming
hydrostatic equilibrium.
We  assume also that the gas temperature is constant 
just inside and outside the density breaks
 (below we discuss the effect of 
a possible temperature gradient in the central region).
 We use the 
temperature values from Fig.~4 
for the external regions and  the 3-D temperature within the group
($T=3.5\pm 0.5$~keV) for the internal region.
The assumption of no temperature gradient on either side of the front
is consistent with our data and is supported by analogy with   A3667
where  the data show
 a clear temperature discontinuity at the edge
 and no temperature gradient  just inside and outside the edge.
Moreover, for A3667 we know that, because the front width 
is  2-3 times smaller than the Coulomb mean free path
 (Vikhlinin et al. 2000a), the two gases at the edge
 are thermally isolated, which we may reasonably expect here as well.

Fig.~5 shows   that the cluster mass 
determinations for  the SE and NW sectors are in
 agreement (as they must be) only at radii larger than the SE edge.
In particular, at the  
 density jump for the SE sector, we find that,
 within the same radius, 
 $M_{int}=(1.9\pm 0.6)\times 10^{13}h_{50}^{-1}
M_{\odot}$ and
$M_{ext}=(9.5\pm 3)\times 10^{13}h_{50}^{-1}M_{\odot}$
(the subscripts $int$ and $ext$ refer to the  estimates from the data
 in the internal and external regions,  respectively; errors are 68\%
confidence level).
Even though the statistics are rather poor,
the mass  appears to be discontinuous, with the mass just
inside the edge  lower than the mass outside.  
This is
obviously an unphysical result.

The discontinuity of the mass estimate comes from the fact that both the temperature
and the density slope ($d\log n/d \log r$) 
decrease as one  crosses the edge from the outside  to the inside.

  Possible explanations for the measured mass discontinuity are:

\begin{enumerate}

\item the temperature within the colder region is not constant; 

\item the gas inside the cloud
is not in hydrostatic equilibrium.

\end {enumerate}

A positive temperature gradient  is not excluded by the
 observed temperature profile (see e.g. Fig.~4).
Such a  gradient would act to increase the mass
 inside the edge, reducing the mass discrepancy.
However, if we make a  simple assumption that within the central core 
 $T\propto r^{\gamma}$, to remove
 the mass discrepancy we need $\gamma\approx 2$.
This implies that the temperature at $r=35^{\second}$ should be $\approx 1.7$ times
 smaller than the temperature at  $r=45^{\second}$ which is clearly inconsistent
with the
observed temperature profile (see Fig.~3 and Fig.~4).

As for the hydrostatic assumption, we know that the 
 subclump moves subsonically relative to the sound speed of the
 external gas where the temperature is $T\approx 10$~keV. However,
 the temperature of the subclump gas is $T\approx 3-4$~keV, which
 means that the
 speed of sound in the gas of the moving subclump is a factor $1.6-1.8$ lower than
 that in the external gas.
Because of  the uncertainties in the measured subclump speed, we cannot 
exclude a supersonic motion of the subclump  relative to 
the sound speed in its own gas.
If it is indeed supersonic, then the  subcluster gas may not have
 sufficient time to react to any perturbations caused by the changing
 external ram pressure, for example, and therefore
 may not be
in hydrostatic 
equilibrium. 
This may distort our mass
 estimate $M_{int}$.     

Regardless of the physical mechanism responsible for  the SE edge mass 
discrepancy, it is clear from Fig.~5 that
 any attempt to measure the total mass using
the hydrostatic equation in circular annuli 
would  result in a significant underestimate of the
total cluster mass at small radii. 

The linear scale of the central
moving gas cloud is approximately equal to the core size for this cluster
(see Table~1) and also to the radial
distance where we expect to find strong lensing 
in clusters like \rxj1720 . 
Moreover, because
 the observed width of the discontinuities is $\le 10^{\second}$,
similar features
may not have been revealed by previous X-ray missions and may,
potentially be responsible for 
some of the  discrepancy 
between the X-ray and the lensing  mass determinations found in some systems
(see e.g. Miralda-Escud\'e \& Babul 1995; but  Allen 1998; and reference therein).


\section{CONCLUSION}
We have presented the results of a  short \chandra ~ observation of the 
cluster of galaxies \rxj1720 . The data indicate a remarkable surface
brightness edge in the SE sector of the cluster  consistent with  
a discontinuity in the density profile along that direction, and a
plateau in the NW sector  consistent with a  break of the gas density 
profile. 
The edge and the break separate the colder inner region from the hotter
external region.
The structure of the SE edge is  similar to the ``cold fronts''  observed
by \chandra ~ in the merging clusters A2142 and A3667. We 
argue that such a discontinuity 
  is produced by the subsonic motion from NW to SE of 
the central cold, group-size  cloud of gas with 
respect to the cluster.
 Unlike A2142 and A3667, \rxj1720 ~ does not show 
the characteristic elongation and 
irregular structure typical of a major merger, and the optical data
show only one central dominant galaxy coincident with
the X-ray surface brightness peak. 
Moreover, the  subclump speed  appears to be small
compared to that expected in the first passage of a subclump infalling from a large distance.
We, therefore, propose a scenario in which \rxj1720 ~
is the result of the collapse of a
group of galaxies  followed by the 
collapse of a much larger, cluster-scale  perturbation at nearly the same
 location in space.
The  available data are consistent with
 this scenario as well as with that of a final
stage of a major merger before the cluster 
becomes fully relaxed.

We also showed that  because of the motion of the central gas cloud,
 the hydrostatic equilibrium equation may underestimate
the true cluster mass in the cluster core.
 Such a phenomenon may explain, in part,
the discrepancy between the X-ray and the strong lensing
mass determinations found in some systems.

\acknowledgements

We acknowledge the entire \chandra ~ team which made this observation
 possible. 
P. M. acknowledge an ESA fellowship and thanks the Center for
 Astrophysics for the hospitality. We thank the referee, C. Sarazin,
 for useful comments and suggestions. 
 Support for this study was provided
 by NASA contract NAS8-39073 and by the Smithsonian
Institution.

\begin{appendix}
\section{DENSITY MODEL}
Below we define the gas density distribution $n_0$ 
that we use to fit the 
surface brightness profile at the SE edge.
We consider a system of spherical coordinates oriented so that
($r,\theta=\pi/2,\varphi=0$) points along the line of sight toward us.
We define a
direction $\hat \delta$ that lies in a plane perpendicular to the
plane of the sky at an angle $\xi$ with respect to the
plane of the sky; this will be the axis of cylindrical symmetry of
our model. The density distribution has three distinct regions
(internal, boundary, and external):

\begin{equation}
n_0(r,\theta,\varphi)\propto \left\{
\begin{array}{*{2}{l}}
 f_{int}(r), & r < r_{int}(\zeta),  \\
&\\
 f_{match}(r), & r_{int}(\zeta)< r < r_{ext}(\zeta),  \\
&\\
 f_{ext}(r), & r < r_{ext}(\zeta),  \\
&\\
\end{array} \right . 
\end{equation}
In the internal and external regions, the profile is:
\begin{equation}
\begin{array}{*{1}{l}}
 f_{int}(r)=A_{jump}\left ({r / r_{jump}} \right)^{-\alpha},  \\
 f_{ext}(r)={\left ( 1+  ( {r_{jump}/  r_{c}} ) ^2 \right )^{3\beta /2}
 /{\left ( 1+  ( {r/ r_{c}} )^2 \right  )^{3\beta /2} }}.
\end{array}  
\end{equation}
These are functions only of the radial distance $r$, and
\begin{equation}
\begin{array}{*{1}{l}}
r_{int}(\zeta)=r_{jump}\exp(-L(\zeta)),\\ 
r_{ext}(\zeta)=r_{jump}\exp(L(\zeta)),
\end{array}  
\end{equation}
are functions only of  the angle $\zeta$ from the direction
$\hat\delta$:
$\zeta(\theta,\varphi)=\sqrt{(\theta-{\pi/ 2})^2+
(\varphi +\xi-{\pi/ 2})^2}$.

Along the direction $\hat \delta$, the width of the boundary layer
 ($r_{int}-r_{ext}$) is
 0 and the density distributions $f_{int}(r)$ and $f_{ext}(r)$ are 
connected at $r=r_{jump}$ by a
density discontinuity  (jump) of amplitude $A_{jump}$. 
For $\zeta>0$, we define a simple 
  ``matching'' function $f_{match}(r)$, which is  a  power-law
that  continuously connects $f_{int}(r_{int}(\zeta))$ with 
 $f_{ext}(r_{ext}(\zeta))$:
\begin{equation}
f_{match}= f_{int}(r_{int})(r/r_{int})^{\nu},
\end{equation}
where
\begin{equation}
\nu=\frac {\log(f_{ext}(r_{ext}))-\log(f_{int}(r_{int}))} {2L(\zeta)}.
\end{equation}

The angular dependence of the width of the boundary layer is given by
equations (A3) and
\begin{equation}
L(\zeta)=S(1-\exp{\left[-{(\zeta^2/ (2 \sigma^2)}\right]}.
\end{equation}

The values $S$ and $\sigma$ can be chosen from a comparison of the
model to 
the observed image. 
We generated a number of models and we found that
  $S=0.5$, $\sigma=40$\degd~ is the best choice.
After these two parameters are chosen from the image, the remaining
parameters, such as the slopes of the density profiles, are derived
from a fit to the brightness profile as described in \S~2.2. 
Fig.~6 shows the corresponding synthesized image in the case 
 $\xi=0$ ( $\hat \delta$ in the
plane of the sky), and the best-fit values from Table~1.

This model appears particularly useful for estimating 
 the angle of the symmetry axis
 $\hat \delta$ with respect to the plane of the sky. 
The shape and the sharpness of the surface brightness profile,
 in fact, are strong functions of
this angle, as shown in  Fig.~7.
Indeed, as the direction $\hat \delta$ moves 
from the plane of the sky, the shape of the surface brightness 
 becomes smother and the edge vanishes.  

We also
ran a number of  simulations  using different matching functions and
we found that, as long as we maintain the condition that the model
image is similar
to the observed one, the important parameters of the density
distribution are not significantly affected.

\end{appendix}

\newpage
\onecolumn


\begin{figure}[h]
\plotone{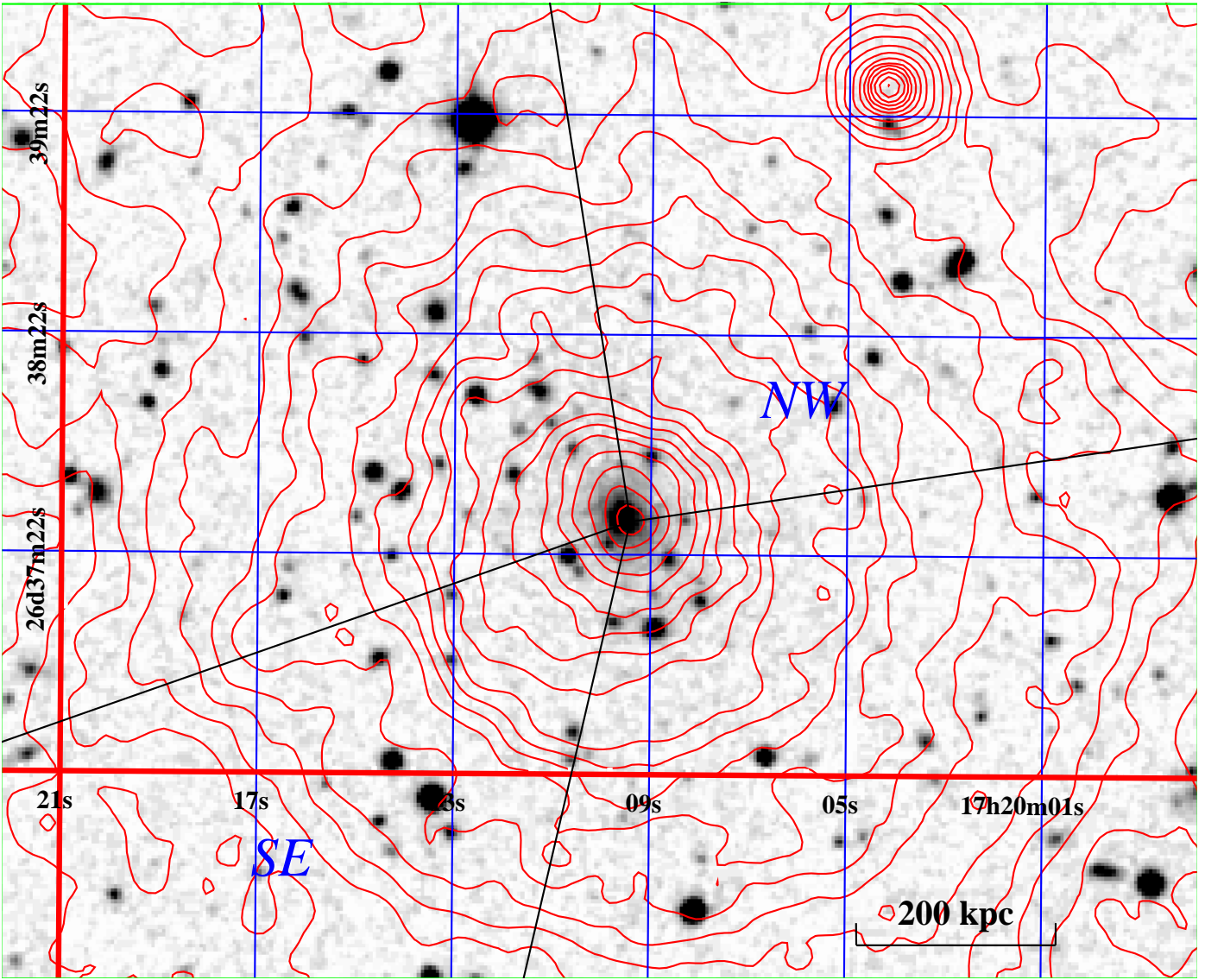}
\caption { {Digitized Sky Survey image with overlaid  ACIS-I X-ray
surface brightness contours (log-spaced by a factor of $\sqrt{2}$) in the
0.5-5~keV energy band after adaptive smoothing. The four straight lines
starting from the X-ray brightness peak  define  
the NW and the SE sectors whose angles are -80\degd~ to 10\degd ~ and
109\degd ~ 169\degd ~   respectively (the position angles are measured
from North toward East). The central galaxy is nearly
coincident with the X-ray peak.}}
\end{figure}


\begin{figure}[h]
\plotone{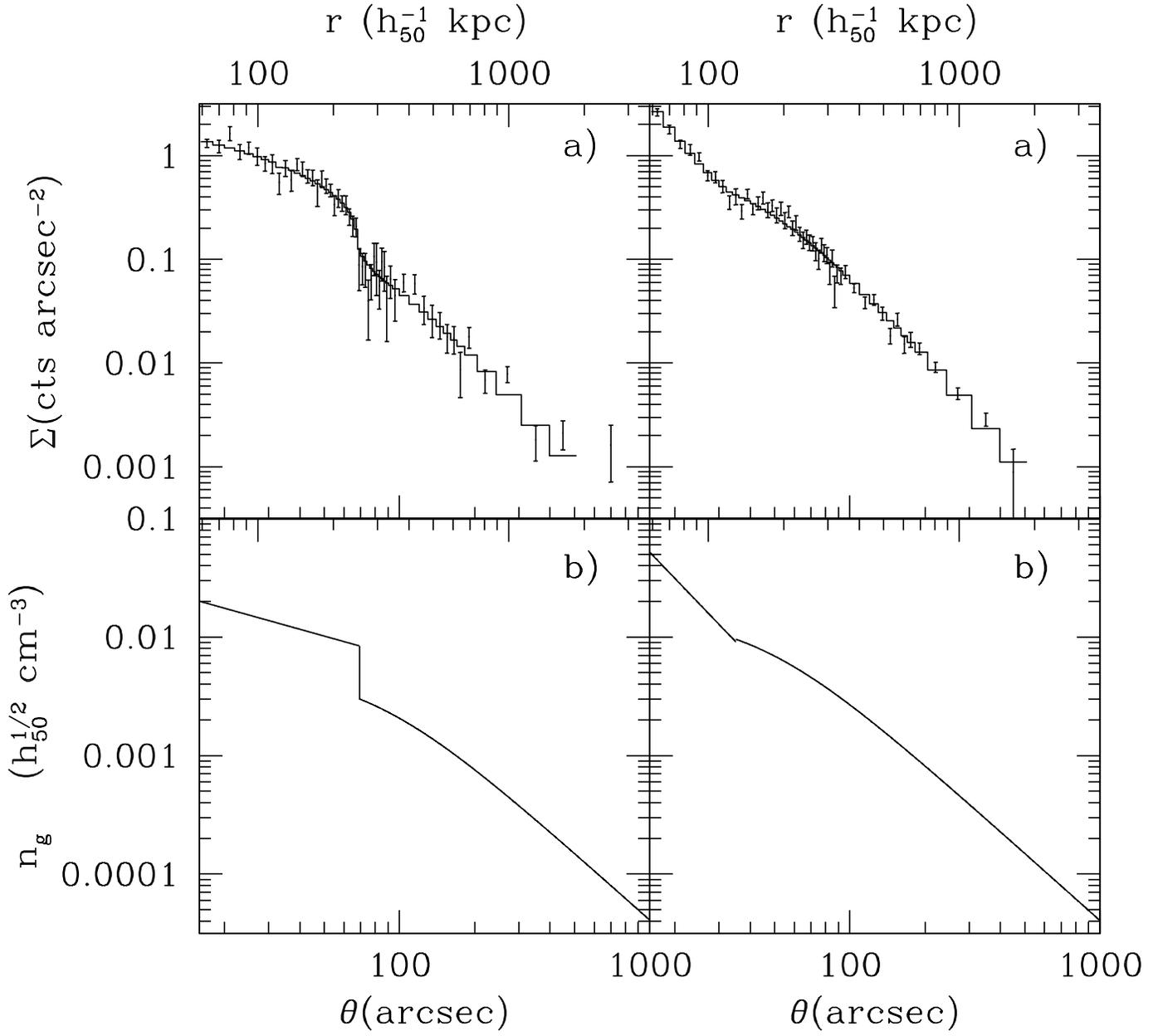} 
\caption {X-ray surface brightness and gas density model 
profiles across the density jumps.  The left
and the right panels refer to the SE and NW sectors respectively.
{\it Panels a)} -- X-ray surface brightness 
profiles across the density jumps. The
 errors are $1\sigma$. The histogram
is the best-fit brightness model that corresponds to the gas density
model shown in panels b). {\it Panels b)} --
 The best-fit gas density model for the X-ray
surface brightness
profile of panels a) (assuming spherical symmetry; see text). }
\end{figure}


\begin{figure}[h]
\plotone{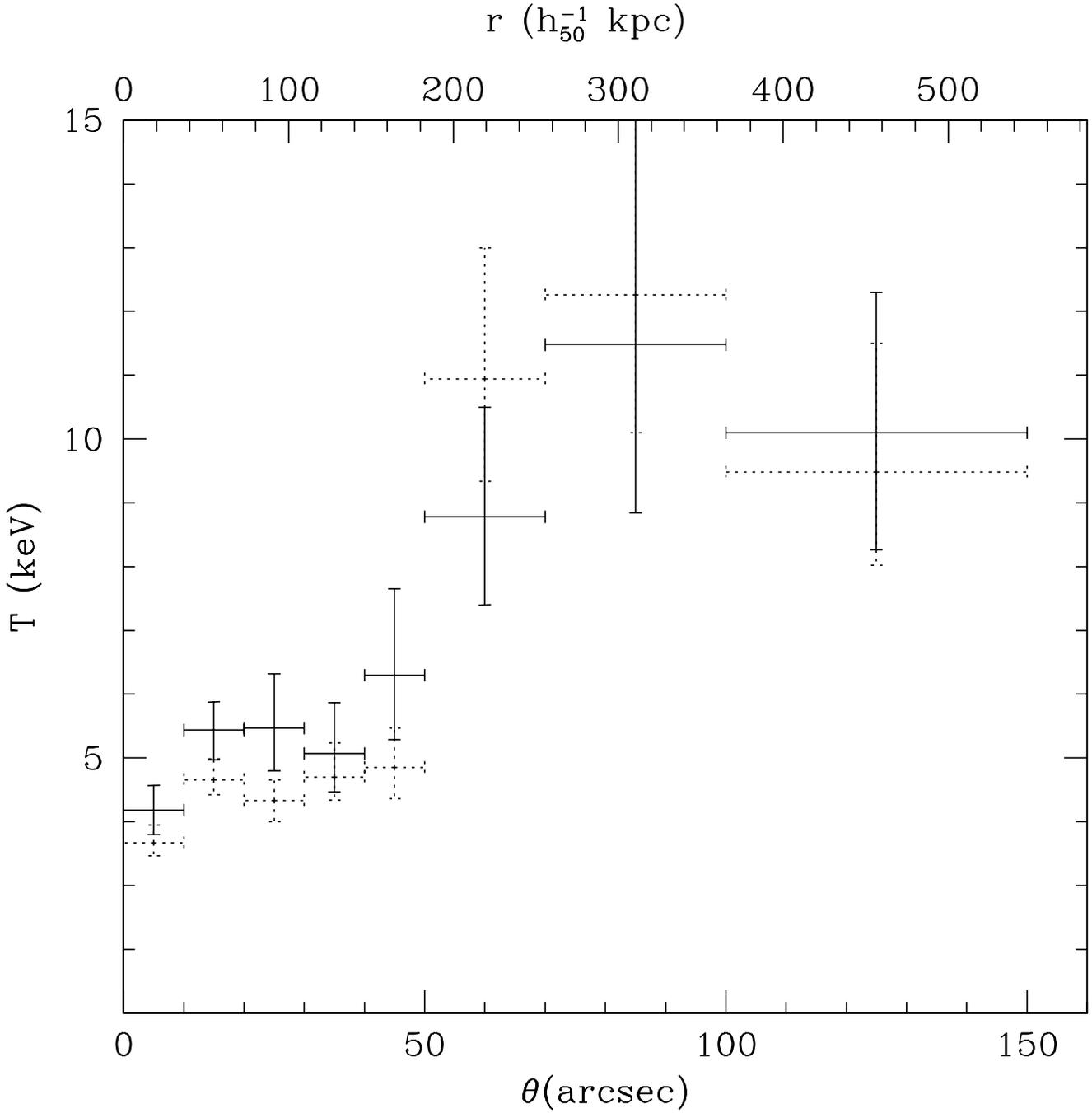} 
\caption { Temperature profile extracted from circular annuli
centered on the X-ray peak (error bars $68\%$ confidence level).Dotted crosses: temperature profile obtained from the
0.6-10~keV band with   $N_H$ as a free parameter.
Solid crosses:  temperature profile in the
1-10~keV band with $N_H$ fixed at the Galactic value.
}
\end{figure}


\begin{figure}[h]
\plotone{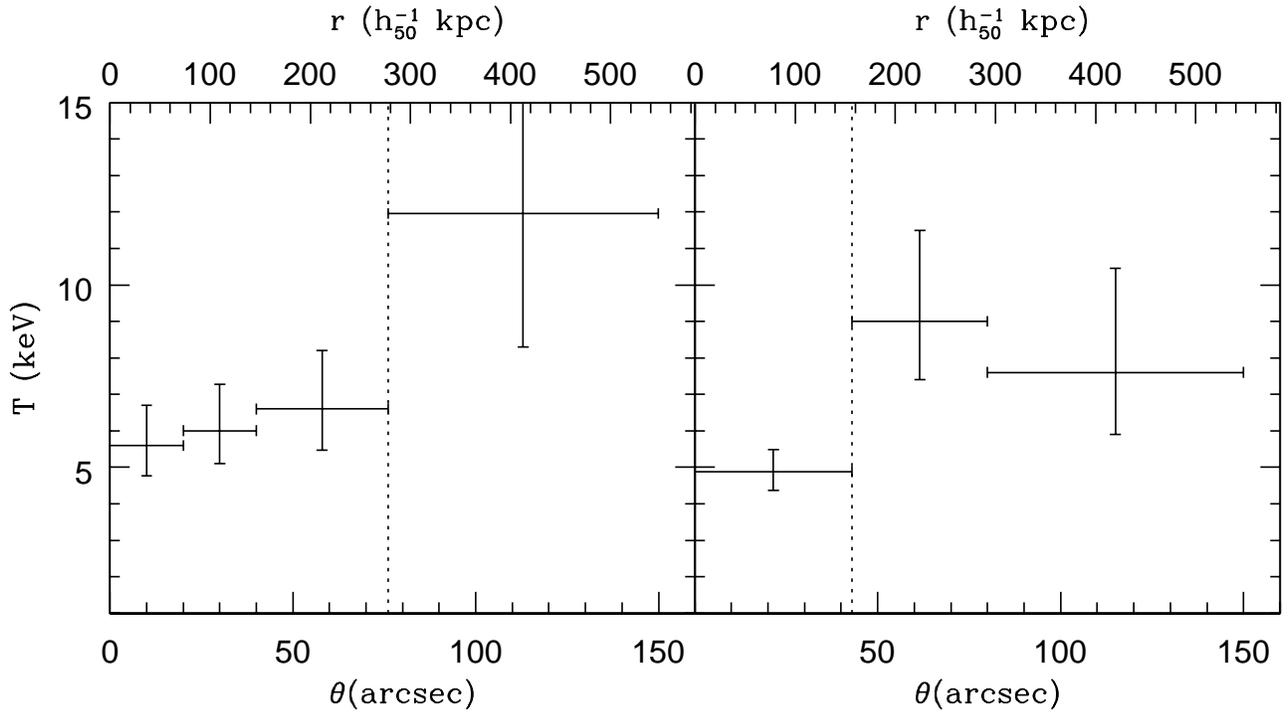}
\caption { Temperature profile extracted from circular annuli
centered on the X-ray  peak in the 1-10~keV band (error bar $68\%$). The left
and the right panels refer to the SE and NW sectors respectively. The
vertical dotted lines indicate the best-fit $r_{jump}$ values.}
\end{figure}


\begin{figure}[h]
\plotone{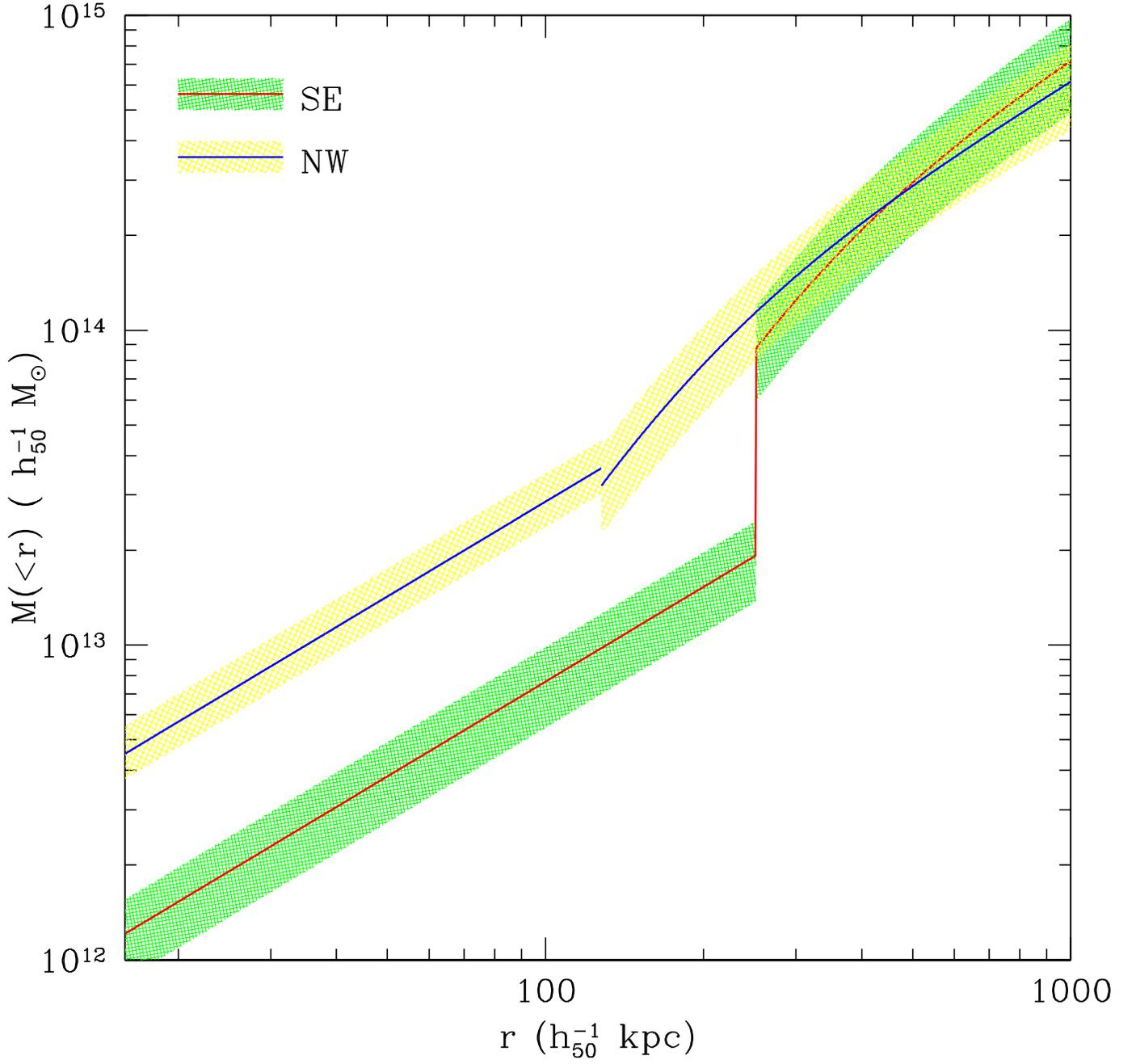}
\caption { The integrated total gravitating mass distribution 
derived, under the assumption of hydrostatic equilibrium, from  
the SE and NW  gas density properties, respectively.}
\end{figure}


\begin{figure}[h]
\plotone{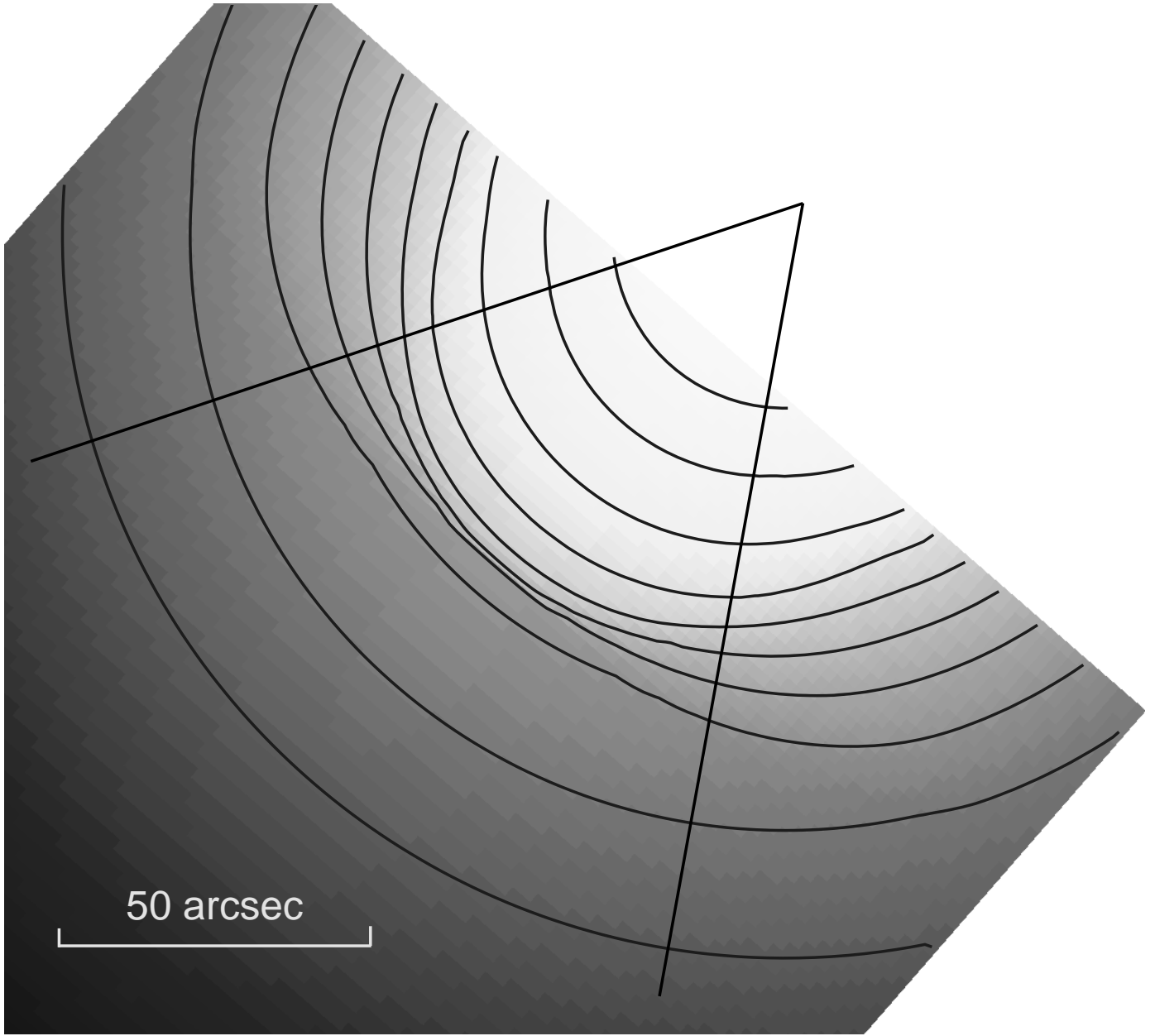}
\caption { X-ray surface brightness image obtained by projecting the
density model defined in the appendix on the plane of the sky.The contours are log-spaced by a 
factor of $\sqrt{2}$. 
Here $S=0.5$, $\sigma=40$\degd~, $\xi=0$\degd~ ( $\hat \delta$ is in the
plane of the sky), and the best-fit parameters are from Table~1.
The two radial lines indicate a sector of angular extent $\psi=\pm
30$\degd .} 
\end{figure}


\begin{figure}[h]
\plotone{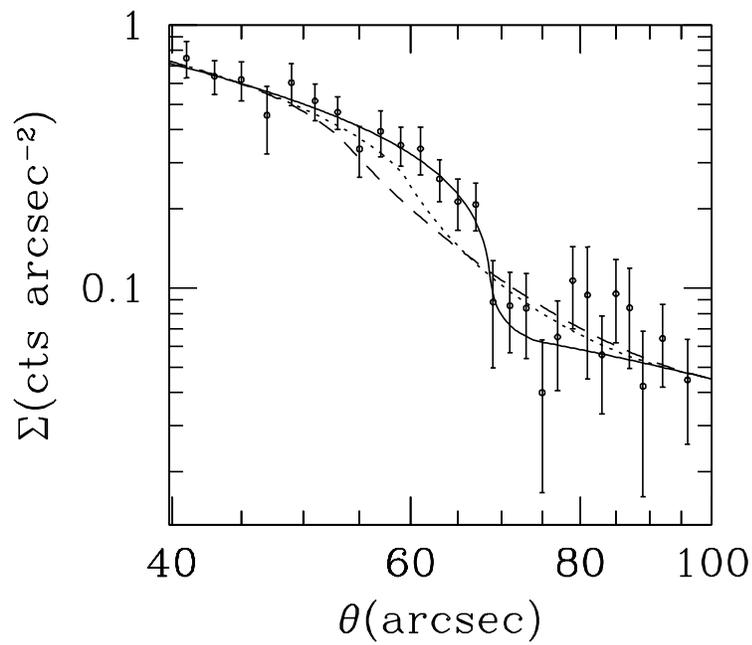}
\caption { Projected surface brightness profiles as a function of the 
direction $\hat \delta$. The density model parameters are the
same as in Fig.~6.
The solid, dotted, and dashed lines 
correspond to
direction angles
$\xi=12$\degd~, $\xi=40$\degd~, and
$\xi=60$\degd~, respectively. 
The data points with error bars correspond to the observed 
SE X-ray surface brightness  
profile (see Fig.~2).}

\end{figure}

\end{document}